\documentclass[prb,aps,twocolumn,groupedaddress,floats,showpacs,final]{revtex4}
\usepackage{graphicx}
\usepackage{dcolumn}
\usepackage{bm}
\usepackage{color}
\usepackage{ulem}
\definecolor{blue}{rgb}{0.3,0.3,0.9}

\def\beq{\begin{eqnarray}}
\def\eeq{\end{eqnarray}}
\def\i{{\rm i}}

\begin{document}

\author{S. Vayl}
\author{A.B. Kuklov}
\author{V. Oganesyan}

\affiliation{Department of Engineering \& Physics,
CSI, CUNY, and The Graduate Center of CUNY}

\title{Sliding phases in U(1) symmetric systems -- mirage of the renormalization group}

\date{\today}
\begin{abstract}
We analyse the proposal of sliding phases (SP) in layers hosting global U(1) symmetric variables with finite inter-layer Josephson coupling.  Based on the Kosterlitz-Thouless  renormalization group (RG) approach, such phases were predicted to exist in various  layered (or 1D quantum coupled) systems. The key in the RG argument is treating the coupling as though the variables are non-compact. Large scale Monte Carlo simulations of a layered model, where the SP is supposed to exist, finds no indication of such a phase. Instead, 3D behavior is observed. This result is consistent with the asymptotically exact analytical solution. A generic argument against SP in translationally invariant  systems with short range interactions is provided. We have also suggested an alternative model for the SP -- adding  long-range interactions to the inter-layer Josephson term.    
\end{abstract}


\maketitle

\section{Introduction}
The idea behind the SP was put forward by Efetov in 1979 in the context of layered superconductor with parallel magnetic field \cite{Efetov}. It was suggested that the field can suppress the inter-layer Josephson coupling so that the low energy properties of this 3D system can be described as being essentially of 2D character. Later, in Ref. \cite{Korshunov}  it was shown that the frustration due to the magnetic field is not sufficient to fully suppress the coupling.

In the context of quantum 1D chains (equivalent to 2D classical layers by the virtue of the quantum to classical mapping) the possibility of the decoupling between chains has also been explored \cite{PWA,Wen}.
The main argument for such a decoupling is based on the scaling dimensions of the Josephson coupling determined with respect to the Luttinger liquid parameter in each chain: if it is larger than 2, the coupling should become irrelevant \cite{Wen}.
These proposals have been criticized in Refs.\cite{Castellani,Fabrizio} where it was shown that the inter-chain tunneling is
always relevant.   

Further argumentation in favor of the SP and, actually, the name {\it Sliding Phases}  have been proposed in Refs.\cite{Lubensky,Toner} where the inter-layer gradient couplings between classical XY variables in each layer have been considered in addition to the Josephson one. Such gradient terms can independently control the scaling dimensions of the Josephson coupling and of the vortex fugacity in each layer so that the first one can become irrelevant above some temperature $T_d$ (of the {\it dimensional reduction} \cite{Sondhi})  which is below the temperature of the Berizinskii-Kosterlitz-Thouless (BKT) transition in the layers. Thus, there is a range of temperatures where the SP are supposed to exist. This approach was also developed for the case of quantum 1D  Luttinger liquids coupled by both the Josephson and the gradient terms \cite{Kane_2001,Ashwin_2001} which are the analog of the Andreev-Bashkin drag effect \cite{AB_effect}.

It is important to note that the proposal of SP  is based on applying the RG logic to compact variables characterized by global U(1) symmetry. While these early suggestions were more of a purely academic interest, expanding capabilities of ultra-cold-atoms techniques in recent years emphasize the importance of these suggestions especially in the context of possible new phases in composite lattices \cite{Cazalila} and in the presence of disorder \cite{Demler}. In more general terms, the question is if it is possible to realize a phase transition from a low- to higher- dimensional behavior.

Here we will analyze a simplest classical  XY system characterized by the gradient interactions and the Josephson coupling $u$. The gradient terms are chosen in such a way that the SP is supposed to exist in some range where the renormalized value $u_r$ of $u$ scales to zero as layers size $L$ grows. We will present the results of the large scale Monte Carlo simulations of this system.  Our analysis is based on the dual formulation of the model  -- in terms of the closed loops. The main result is that no SP state exists in such  a system. Instead, the value of $u_r$ is always finite. This behavior will be compared with the standard  asymmetric XY layered model where no SP are expected to occur.
We will also derive the analytical result for $u_r$ in the asymptotic  limit when the intra-layer stiffness is much larger than $u$. 
The numerical results have been found to be consistent with the analytical solutions for both models. 

Our paper is organized as follows. In Sec.\ref{Sec:I} we introduce the bilayer model and provide the RG solution for SP. Then, we construct the dual representation in Sec. \ref{sec:dual}. Using the duality we have found the asymptotic analytical solution for the renormalized Josephson coupling $u_r$ in Sec. \ref{sec:AS}. The Monte Carlo simulations of the bilayer model are presented in Sec.\ref{sec:num}. Then, in Sec.\ref{Sec:Nz} we present the results on a stack of bilayers along the same lines as for the bilayer. Finally, in  Sec. \ref{sec:dis} we discuss the implications of our analytical and numerical results and also provide an alternative model for the SP. 
 
\section{Bilayer model of SP}\label{Sec:I}
Here we introduce a model of two asymmetric parallel layers, each being a square lattice of linear size $L=1,2,3,...$ (in terms of the inter-site shortest distance) characterized by two fields $\psi_1=\exp(\phi_{1})$ and $\psi_2=\exp(\phi_{2})$ on the layers $z=1,2$, respectively.The action can be written as
\beq
H&=& - \sum_{\langle ij\rangle} [t_1 \cos(\nabla_{ij} \phi_1 - A_{ij}) +t_2 \cos(\nabla_{ij} \phi_2 -g_2 A_{ij}) 
\nonumber \\
 &+& \frac{1}{2g} A^2_{ij}] - \sum_i u\cos(\phi_2(i)-\phi_1(i))
\label{2N} 
\eeq
where $t_1>,t_2>0, g>0$ and $g_2$ are parameters; $\langle ij\rangle$ denotes summation over nearest neighbor sites within each layer; $\nabla_{ij} \phi_a \equiv \phi_a(i) - \phi_a(j)$; $A_{ij}$ is a bond vector field (that is, $A_{ij}=-A_{ji}$) oriented along the bond $\langle ij\rangle$. It is introduced in order to generate the "current-current" interaction (cf. \cite{Lubensky,Toner,Sondhi,Kane_2001, Ashwin_2001}) consistent with the compact nature of the fields $\phi_{1,2}$. This action is to be used in the partition function 
\beq
Z=\int DA D\phi_{1} D\phi_{2} \exp(- H)
\label{ZZ2}
\eeq
 where the temperature is absorbed into the the parameters $t_1,t_2,u,g$. Our focus is on verifying the applicability of the RG analysis to the renormalization of the Josephson coupling $u$. Hence, we will not discuss physical origins of the variables and the parameters.

\subsection{The RG solution for bilayer}\label{sec:RG}
In  the approximation ignoring compact nature of the variables, the terms $-\cos(\nabla_{ij} \phi_1 - A_{ij})$ and $-\cos(\nabla_{ij} \phi_2 -g_2 A_{ij})$ are replaced by
$(\nabla_{ij} \phi_1 - A_{ij})^2/2$ and $(\nabla_{ij} \phi_2 -g_2 A_{ij})^2/2$, respectively. Then, the gaussian integration over $A_{ij}$ can be carried out explicitly in Eq.(\ref{ZZ2}), so that (\ref{2N}) in terms of the remaining variables becomes
\beq
H_0= \frac{1}{2}\sum_{\langle ij\rangle} K_{ab} \nabla_{ij} \phi_a \nabla_{ij} \phi_b - \sum_i u\cos(\phi_2-\phi_1),
\label{2N0} 
\eeq
 where the matrix $K_{ab}, \, a,b =1,2$ is related to the original parameters as
\beq
K_{11}&=& \frac{t_1(1+ gg_2^2t_2)}{1+g(t_1 +g_2^2t_2)},\, K_{22}= \frac{t_2(1+ gt_1)}{1+g(t_1 +g_2^2t_2)},
\nonumber \\
 K_{12} &=& - \frac{g g_2 t_1 t_2}{1+g(t_1 +g_2^2t_2)}.
\label{Ktg}
\eeq
[As a matter of taste, we will keep $g_2<0$ in order to have $K_{12} >0$].
 The stability of $H_0$ is guaranteed by 
\beq
K_{11}K_{22} - K_{12}^2=\frac{ t_1 t_2}{1+g(t_1+g_2^2t_2)} >0. 
\label{stab}
\eeq
The condition for SP can be obtained along the lines of the logic \cite{Wen, Lubensky,Toner,Sondhi,Kane_2001, Ashwin_2001} which ignores the compactness of $\phi_{1,2}$. Specifically, introducing the variables $\varphi =\phi_1 + \phi_2$ and $\theta= \phi_2 -\phi_1$ and, then, integrating out $\varphi$, the resulting partition function becomes 
\beq
Z_0=\int D\theta {\rm e}^{-H_\theta},\,  H_{\theta} =  \int d^2 x  \left [ \frac{K}{2} (\vec{\nabla} \theta)^2 - u \cos\theta\right],
\label{Z_0} 
\eeq
where the notation
\beq
K=\frac{K_{11}K_{22}-K_{12}^2}{K_{11}+K_{22}+ 2K_{12}}
\label{K}
\eeq
is introduced and the long wave limit is considered -- so that the summation along the layers is replaced by the integration $\int d^2 x ...$. 

As long as the compactness of $\theta$ is ignored, Eqs. (\ref{Z_0}) represent the standard Sine-Gordon model in 2D. The RG analysis predicts (see in, e.g., \cite{Lubensky_book}) that at $K<K_d= 1/(8\pi)$ the renormalization renders the Josephson coupling $u$ irrelevant in the thermo-limit $L\to \infty$. More specifically, the renormalized $u$ should flow to zero as $u_r \sim u L^{b}\to 0,\, b= 2(1- K_d/K)<0$. Such a behavior is supposed to occur together with the persistence of the algebraic order along the planes. This requirement imposes further restrictions on the values of $K_{ab}$. 

Without loss of generality let's assume $K_{11} <K_{22}$ and introduce the notations:  $T=1/K_{11}$ as a measure of temperature, and
 $Y=K_{22}/K_{11} >1,\, X=K_{12}/K_{11} $. Then, the condition $K<1/(8\pi)$ for SP becomes
\beq
T>T_d=\frac{8\pi (Y-X^2)}{1+Y + 2X}.
\label{SSP}
\eeq

In order to guarantee the algebraic order in each layer no BKT transition should occur in the layers.  In order to determine possible types of vortices responsible for the transition, we examine the form (\ref{2N0}) by "reinstating" the compactness of the variables
in the limit $u=0$ (which is supposed to renormalize to zero). In the presence of the gradient coupling a vortex may have a composite structure \cite{Babaev,Kaurov}: $q_1$ circulations in the component 1 may be bound to $q_2$ circulations in the component 2. Finding the condition for the proliferation of such composite vortices can be achieved along the line of Kosterlitz-Thouless argument developed for simple XY-model in 2D. The free energy of such a complex is
\beq
F_v= \pi [K_{11} q_1^2 + K_{22} q_2^2 + 2K_{12} q_1 q_2] \ln L  -2 \ln L. 
\label{vort}
\eeq
Then, $F_v\propto  \pi [(q_1+ Xq_2)^2 + (Y-X^2)q^2_2]  - 2T $, and the stability against the BKT transition is guaranteed by the positivity of $F_v$ or
\beq
T < T_{(q_1,q_2)}= \frac{\pi}{2}[(q_1+ Xq_2)^2 + (Y-X^2)q^2_2],
\label{TBKT}
\eeq 
where the minimization with respect to $q_1, q_2$ must be performed. Proliferation of simple vortices $q_1=\pm 1,\,\, q_2=0$ corresponds to
$ T_{(1,0)}=\pi/2$. It is, however, not always a minimal value as long as $X\neq 0$. Let's also note that, since $Y>1$ by definition, there are no 
solutions for $T_d < T_{(q_1,q_2)}$ if $X=0$. 

Let's look for a solution when $X$ is integer, that is, $X=1,2,3,...$. Then, keeping in mind the stability requirement (\ref{stab}), that is, $Y-X^2=\delta >0$,
the minimal vortex corresponding to $q_1 =- Xq_2, q_2=\pm 1$ has free energy lower than that of the simple vortex if $\delta <1$. 
In this case, the solution for $T_d <  T_{(X,-1)}$ exists if $X\geq 3$. 
Proliferation of the composite vortex pairs corresponds to disordering of the original fields $\exp(i\phi_{1,2})$, while the composite field $\Psi= \exp(i (\phi_1 + X \phi_2))$ remains (algebraically) ordered. This mechanism constitutes the formation of thermally induced bound phases (or using the language of superfluidity -- {\it Thermally Paired Superfluid}, TPS,  \cite{TPS}). In other words, the system behaves in a such a way that the algebraic order persists only in the composite field $\Psi=\psi_1 (\psi_2)^X$.[Since $X>1$ we call such a composite phase as thermally bound superfluid (TBS) by analogy with the TPS]. This effect does not require that $X$ is necessarily integer. If $X$ is non-integer, its closest integer part will determine the power of $\psi_2$. 

If $\delta >1$, the lowest energy vortex is the simple one, and the solution for the inequality $T_d < T_{(q_1,q_2)}$ also requires $X\geq 3$. 
For $X>>1$, $T_d \to 0$ while $ T_{(q_1,q_2)} \to (\pi/2) {\rm min}(1,\delta)$ as long as
$\delta $ is kept constant. Such a limit corresponds to the largest range of $T$ where SP are to be anticipated for the two-layer model. 
However, for practical purposes of simulations using too large $X$ leads to slower convergence. Thus, we choose $X=5,\, Y=25.5$ corresponding to a reasonably wide range where SP is anticipated to exist.
Taking into account Eq.(\ref{K}), the condition for the SP can be written as
\beq
\frac{8\pi \delta }{\delta +(1+X)^2} < T < \frac{\pi}{2} \delta.
\label{SPT}
\eeq
 Keeping in mind the chosen values $X=5, \delta=1/2$, this becomes $8\pi/73 < T < \pi /4$ or $ 0.344< T<0.785$. The results of the simulations will be conducted at the "temperature" $T$ in the middle of the interval $(T_d,T_{BKT})$, that is, $T\approx 0.565$. More specifically, $K_{11}=1/T,\, K_{22}= 25.5 K_{11},\, K_{12}=5K_{11}$. In terms of the original parameters $t_1, t_2, g_2, g$, the relations
$gt_2|g_2|(1-5|g_2|)=5$, $gt_1(5.1|g_2|-1)=1$ , $t_1\approx 0.177 |g_2|/[(1-4|g_2|)(5.1|g_2|-1)$ and $10/51 < |g_2| <1/5$ must be satisfied. [Such a fine tuning is of no concern, since the focus is on the consistency of the paradigm of SP rather than on physical origin of the values].

\begin{figure}
\vspace*{-0.5cm}
 \includegraphics[width=1.0 \columnwidth]{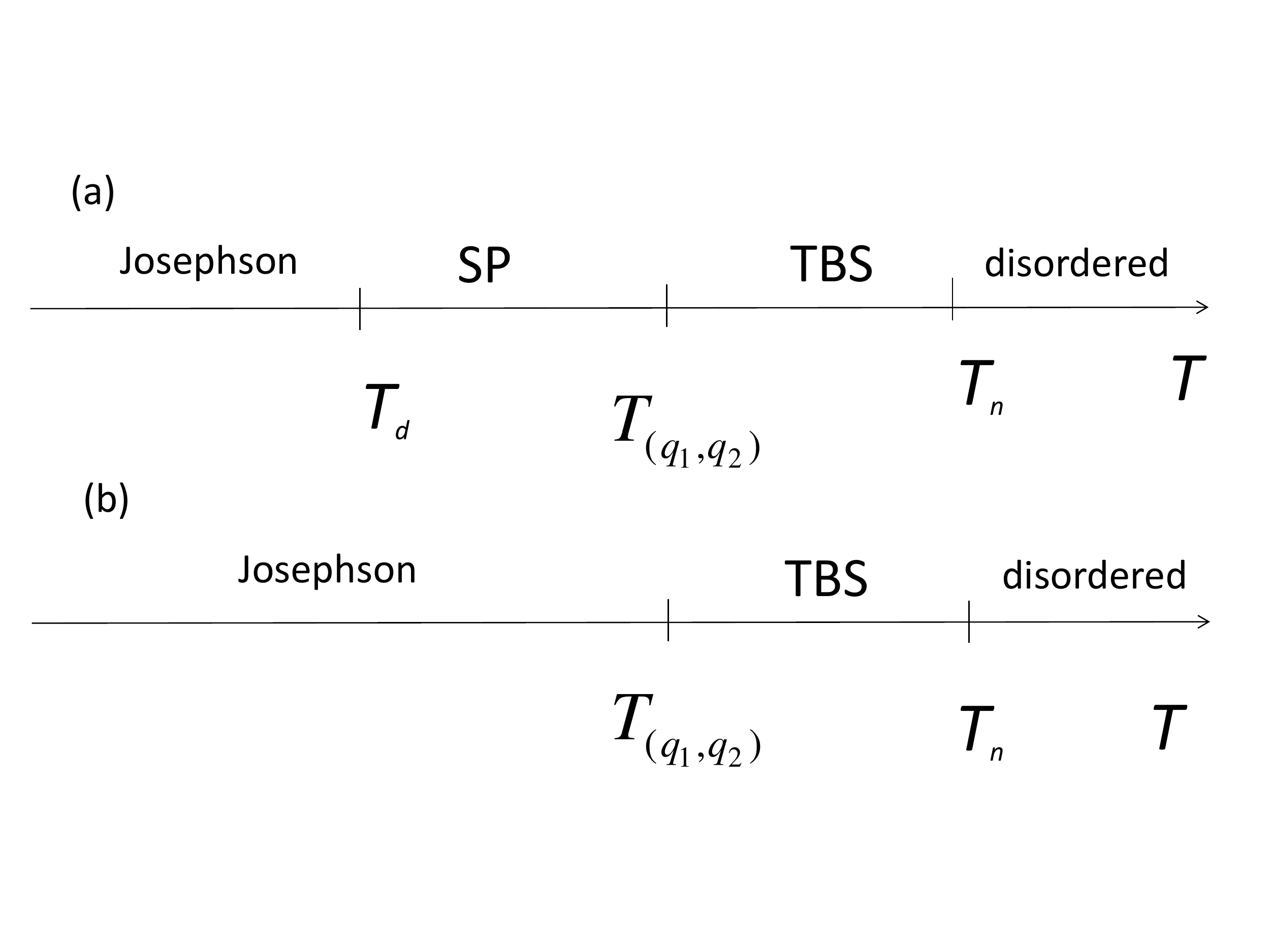}
\vskip-8mm
\caption{(Color online)  Two options for phases in the bilayer model: (a) with the SP according to RG; (b) Without SP. }
\label{PD}
\vskip-5mm
\end{figure}

Concluding this section, the presented analysis based on the RG finds that it is possible to find a range of temperatures where the sequence of phases is as presented in Fig.~\ref{PD} in the panel (a): at $T<T_d$ the Josephson  coupling is relevant. We call this range as Josephson phase in Fig.~\ref{PD}. At $ T_d <T< T_{(q_1,q_2)}$ there is the SP where the symmetry $U(1)$ is promoted to U(1)$\times$ U(1). In the range $ T_{(q_1,q_2)}<T<T_n$ the TBS phase is characterized by the composite field $\Psi$.  Thus, the broken symmetry is partially restored through the subgroup $Z_N$, where $N=1+[X]$. At higher temperatures $T>T_n$ the composite field $\Psi$  becomes disordered.  

\subsection{Dual representation}\label{sec:dual}    
The partition function $Z$
can be evaluated by the high-temperature expansion method (see e.g. in \cite{Parisi}) in terms of $t_1,t_2,u$ and the explicit integration over the variables. This approach allows obtaining $Z$ in terms of the integer bond variables -- powers of the corresponding Taylor series. Since the resulting configurational space consists of closed loops of the bond currents, further simulations can be effectively performed by the Worm Algorithm \cite{WA}. As will be also shown, the language of loops also allows obtaining analytic expressions for the renormalized Josephson coupling $u_r$ which are exact in the asymptotic limit.  

We will be utilizing the Villain approximation \cite{Villain} for the cosines to obtain the so called J-current version \cite{Jcurr} of Eqs.(\ref{ZZ2}),(\ref{2N}):
\beq
Z=\sum_{\{m_{a,ij}, m_i\}} \int D\phi \int DA {\rm e}^{-H_V}, 
\label{Vill}
\eeq
\beq
H_V&=& \sum_{\langle ij\rangle} [\frac{\tilde{t}_1}{2} (\nabla_{ij} \phi_1 - A_{ij} +2\pi m_{1,ij})^2 
\nonumber \\
&+&\frac{\tilde{t}_2}{2} (\nabla_{ij} \phi_2 -g_2 A_{ij} +2\pi m_{1,ij} )^2 +\frac{1}{2g} A^2_{ij}]
\nonumber \\
 &+&  \sum_i \frac{u_V}{2} (\phi_2(i)-\phi_1(i) +2\pi m_i)^2,
\label{2Nv} 
\eeq
where $m_{a,ij}= - m_{a,ji}=0,\pm 1, \pm 2,...$ ($a=1,2$) are integer numbers defined along bonds between two nearest sites $i$ and $j$ along the planes and $m_i=0,\pm 1, \pm2, ...$ is an integer assigned to a site $i$ and oriented from the layer 1 to the layer 2. 

  The Villain approximation proves to be very accurate  for establishing the transition points as well as in general if the effective constants $\tilde{t}_1,\tilde{t}_2, u_V$ are properly expressed in terms of the corresponding bare values $t_1,t_2,u$ (see in Ref.\cite{Kleinert}).
The "renormalization" can be essentially ignored for $t_1,t_2 \geq 1$, so that in what follows we will be using $\tilde{t}_1=t_1,\, \tilde{t}_2=t_2$. Similarly, for the Josephson coupling $u \sim 1$ one should take $u_V=u$ and, if $u<<1$, the corresponding relation is $u_V=1/(2 \ln (2/u))$ \cite{Villain,Kleinert}. 
  
  After using the Poisson identity for each integer and performing the integrations over $\phi_i$ and $A$, the resulting expression becomes
\beq
Z=\sum_{\{J_{1,ij}\},\{J_{2,ij}\}, \{J_{z,i}\}} {\rm e}^{-H_J},
\label{Z2}
\eeq
\beq
H_J= \sum_{\langle ij \rangle; a,b} \frac{1}{2}(K^{-1})_{ab} J_{a,ij} J_{b,ij} + \sum_i \frac{1}{2u_V} J_i^2,    
\label{H_J}
\eeq
where $a,b=1,2$ labels layers and $(K^{-1})_{ab}$ is the matrix inverse to $K_{ab}$ introduced in Eqs.(\ref{2N0}),(\ref{Ktg}); The summation runs over the integer bond currents $ J_{1,ij}, \, J_{2,ij}$  (oriented from site $i$ to site $j$ so that $ J_{a,ij}=-  J_{a,ji},\, a=1,2$) within each corresponding layer 1,2 as well as over the integer currents $J_{i}$ oriented along the bond connecting the site $i$ in the layer 1 to the site $i$ in the layer 2. All the configurations are restricted by the Kirchhoff's current conservation rule -- the total of all J-currents  incoming to any site must be equal to the total of all outcoming currents from the same site. 

It is useful to note that Eqs.(\ref{Z2},\ref{H_J}) can also be obtained directly from Eq.(\ref{2N0}) by reinstating the compact nature
of the variables: $ \nabla_{ij} \phi_1 \to \nabla_{ij} \phi_1 +2\pi m_{1, ij}$ and $ \nabla_{ij} \phi_2 \to \nabla_{ij} \phi_2 +2\pi m_{2, ij}$ so that the matrix $K_{ab} $ is viewed as being independent from the parameters in the action (\ref{2N}).

The system (\ref{Z2},\ref{H_J}) features statistics of closed loops. If $u=0$, there are two sorts of loops -- one in each layer. 
Thus, each configuration is characterized by definite values of the windings $W_{a, \alpha}$ in the $a$th layer along the $\alpha=\hat{x},\hat{y}$ directions of the planes. It is straightforward to show that statistics of these windings determine the renormalized values $\tilde{K}_{ab}$ of the matrix $K_{ab}$ along the line of the approach \cite{Ceperley}. More specifically
\beq
\tilde{K}_{ab} = \frac{1}{2} \sum_{\alpha=\hat{x},\hat{y}} \langle W_{a,\alpha} W_{b,\alpha}\rangle.
\label{KR}
\eeq  
This expressions are valid for periodic boundary conditions (PBC).
It is important to note that $\tilde{K}_{ab}$ represents an exact linear response (at zero momentum) with respect to Thouless phase twists.
In other words, if there are externally imposed infinitesimal constant gradients $\nabla_\alpha \phi_{1,2} \to 0$ (violating the PBC) of the phases $\phi_{1,2}$,   the free energy acquires the contribution $\delta F=\frac{1}{2}L^2 \sum_{a,b, \alpha} \tilde{K}_{ab}\nabla_\alpha \phi_{a}\nabla_\alpha \phi_{b} $.  On the other hand, in the presence of the gradient the integrand of the partition function gets the factor $\exp(i L\sum_{a,\alpha} W_{a,\alpha} \nabla_\alpha \phi_a)$. Comparing both expressions leads to the relation (\ref{KR}).

As a test of consistency, we have checked numerically that in the regime where the SP state is supposed to exist (that is, $X=5,Y=25.5, T\approx 0.565$), the deviations of $\tilde{K}_{ab}$ from the bare values $K_{ab}$ are within the statistical error less than 1\% for all tested sizes of the layers $10\leq L \leq 1000$. Significant deviations are observed only as the system approaches fully disordered state -- that is, where the fields $\psi_{1,2}$ as well as the composite one $\Psi$ become disordered. In this case, $\tilde{K}_{ab}$ flow to zero as $L$ increases. The deviations remain small (about 2-3\%) even in the regime where $\Psi$ is the only ordered field. The emergence of the TBS is detected by observing that windings $W_{a,\alpha}$ in the layers 1 and 2 are changing exactly by the increment $\Delta W_1= 1$, $\Delta W_2= X$ (plus or minus), respectively.   
 
At finite values of $u$ the loops belong to both layers so that no separate windings can be introduced. However, the sums $W_\alpha = W_{1,\alpha} +  W_{2,\alpha}$ remain well defined and can be used to evaluate the rigidity $\rho_{\alpha}$ of the fields along the layers. In a general case of $N_z$ symmetric (with respect to the $x,y$ directions) layers $\rho=\rho_x =\rho_y$ :
\beq
\rho&=&\frac{1}{2N_z}\sum_\alpha \langle W^2_\alpha\rangle     
\label{stif22} \\
W_{\alpha} &=& \frac{1}{L} \sum_{\langle ij \rangle, a=1,2,...N_z} J_{a,ij},
\label{WW}
\eeq
where for a given $\alpha=\hat{x},\hat{y}$ in (\ref{WW}) the bond $\langle ij \rangle$ (as well as $J_{a,ij}$) is oriented along the direction $\alpha$.

Our focus here on the renormalized value $u_r$ of the Josephson coupling $u$ in the SP regime. If the periodic boundary conditions are also imposed perpendicular to the layers (along $z$-direction), the inter-layer response $u_r$ is given by windings $W_z$ along $z$-direction:
\beq
u_r=\frac{N_z}{L^2}\langle W^2_z\rangle , \quad       W_z = \frac{1}{N_z} \sum_{i} J_i,
\label{WU}
\eeq
where the summation $\sum_i$ of the currents $J_i$ (oriented along $z$-direction) is performed over all sites of all layers.
 Similarly to the cases (\ref{KR}) and (\ref{stif22}), Eq.(\ref{WU}) represents the full linear response at zero momentum -- that is, the renormalized value $u_r$ of the Josephson coupling $u$. 

At this point, we should comment on how to interpret the PBC for two layers, $N_z=2$. While in the case $N_z\geq 3$ it is a natural procedure to link the $z=N_z$th layer to the first one, $z=1$, by the Josephson  term, the case $N_z=2$ needs an auxiliary  construction because the layers 1 and 2 are coupled already directly. The formal procedure, then, consists of adding a third layer, $z=3$, with no rigidity along $x,y$ directions and coupled by the Josephson term to both layers, $z=1,2$. If the coupling $u_{13}$ between the layers 1 and 3 and the coupling $u_{23}$ between the layers 2 and 3 add up as $1/u_{13} + 1/u_{23}=1/u_V$, in the dual action (\ref{H_J}) the sum in the last term can be extended to the layers $z=1,2,3$, while the first term is still confined to the layers $z=1,2$. The key to this procedure is the Kirchhoff's rule: the J-current from a site $(x,y)$ along $z$-direction from the layer 2 to the layer 3 must be exactly the same as the current from the site $(x,y)$ in the layer 3 to the layer 1.  Then, in the form (\ref{H_J})  the same value $u_V$ can be used for the currents from the layer 1 to the layer 2 directly  or through the layer 3.      

\subsection{Asymptotic expression for $u_r$}\label{sec:AS}
As mentioned above, the dual representation allows obtaining analytically the asymptotic values for $u_r$ . Let's begin with the trivial case of zero stiffnesses $K_{ab}$ and arbitrary number of layers, $N_z=2,3,4,$. In this case, the action (\ref{H_J}) trivially becomes
\beq
H_A= \frac{N_z}{2u_V} \sum_{i} J^2_i, \quad J_i=0,\pm 1, \pm 2,...
\label{A}
\eeq
where the summation runs over all sites $i$ of only {\it one} layer, say, $z=1$. In this expression the Kirchhoff rule dictates that the current
$J_i$ at a given site along $z$-direction must be the same for all values of $z$. Thus, such a current with $J_i=\pm 1$ constitutes one closed loop characterized by the winding $W=J_i$. This allows constructing the partition function exactly as 
\beq
Z_A=\left[\sum_{W=0,\pm 1, \pm 2, ...} \exp\left(-\frac{N_z}{2u_V}W^2\right)\right]^{L^2}
\label{ZA}
\eeq
where $L^2$ is the number of sites in one layer. The stiffness (\ref{WU}) can be found by taking into account that the total winding along z-direction is $W_z=\sum_i J_i$, where the summation runs over $L^2$ sites of only one layer. Then, using statistical independence of different sites we find
\beq
u_r=\frac{2N_z\sum_{W=1, 2,...} W^2\exp\left(-N_zW^2/2u_V\right)}{1+2\sum_{W= 1, 2,...} \exp\left(-N_zW^2/2u_V\right) }.
\label{UR1}
\eeq
This expression shows that, as long as $N_z$ is finite, the Josephson coupling remains relevant even if there is no in-plane order.
  
In the limit $u_V <<1$ only the term $W=1$ is important, so that Eq.(\ref{UR1}) becomes
\beq
u_r= \frac{2N_z}{2+\exp(N_z/2u_V)}.
\label{URR}
\eeq 

 Eq.(\ref{URR}) can be used even in the case when there is finite stiffness along the layers, with $N_z$ substituted by some effective value $M$, that is, 
\beq
u_r= \frac{2M}{2+\exp(M/2u_V)}. 
\label{GenM}
\eeq
This is easily justified in the "ideal gas" approximation of rare fluctuations of the J-currents in z-direction. In the case of  $K_{12}=0$ the asymptotic asymmetric limit corresponds to $K_{11}=K_{22} >> 1$ (so that the system is well above the disordering transition)  and $u_V<<1$. 
The loop proliferation can be viewed from the perspective of the Worm Algorithm \cite{WA} where one open end of a string of J-currents walks randomly  until it meets another open end so that the loop is formed. Then, the most of the path is residing in a layer with only occasional jumps between neighboring layers. These jumps are controlled by the exponential smallness of $\exp(-1/2u_V)$. Thus, the full action can be well approximated by $H_A$, Eq.(\ref{A}), with $N_z=1$. This leads to Eq.(\ref{GenM}) with $M=1$: 
\beq
u_r= \frac{2}{2+\exp(1/2u_V)}.
\label{urXY}
\eeq 
 Later we will present the numerical evidence that the 
stiffness perpendicular to the layers of a simple XY layered model in the asymptotic limit can well be described by the above equation.
Below we will show that for the case of the two asymmetric layers, the effective value of $M$ is $M=2$ in Eq.(\ref{GenM}).

\subsection{Numerical results for $N_z=2$}\label{sec:num}
Here we present the results of Monte Carlo simulations of the bilayer in the regime of SP in the limit $u_V<<1$.
The action (\ref{H_J}) can be represented in the notations $T,X,Y,\delta$ as
\beq
H_J= \sum_{\langle ij\rangle}\left[ \frac{T}{2} J_{1,ij}^2 + \frac{T}{2\delta}(J_{2,ij}-XJ_{1,ij})^2\right] + \sum_i \frac{J^2_i}{2u_V},
\label{H12}
\eeq
where $J_{1,ij}$ and $J_{2,ij}$ refer to the inplane bond currents in the layers 1 and 2, respectively; the actual values of the parameters  used in the simulations have been discussed at the end of Sec.\ref{sec:RG}: $X=5, \delta=1/2, T=(T_d + T_{(X,-1)})/2 \approx 0.565$. 

\begin{figure}
\vspace*{-0.5cm}
 \includegraphics[width=1.1 \columnwidth]{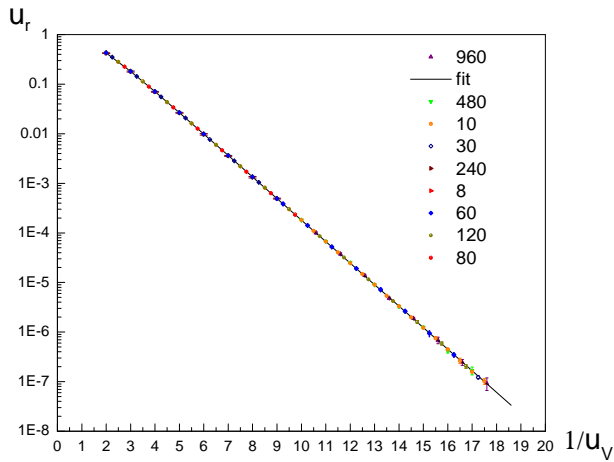}
\vskip-8mm
\caption{(Color online) Monte Carlo results for the inter-layer stiffness $u_r$ vs its bare value $u_V$ of two layers for various layer sizes (shown in the legend).Error bars are shown, and for the majority of the data points these are smaller than symbols. The fit line is the solution (\ref{UR3}).  }
\label{figN2}
\vskip-5mm
\end{figure}

The structure of the loops is determined by the energy of creating a J-current element along a given direction. A typical energy to create an additional J-current element in the plane 2 can be estimated as $ \delta E_2 \approx T/(2\delta) \approx 0.5$.
Thus,  large loops with a typical values $|\vec{J}_2|=1$ can exist in the plane 2. In contrast, the energy to create an isolated element in the plane 1 (with no $J_2$ currents along the same bond in the layer 2)  costs much larger energy: $\delta E_1 \approx T(1+ X^2/\delta)/2 \approx 15 $ . Thus, the probability to create such an element is exponentially suppressed as $\sim \exp(-15)$, and no entropy contribution (due to 4 optional directions along the plane) can compensate for such a low value. This implies that no large isolated loops can exist in the layer 1. The only option to create a large loop in the layer 1 is if each element $J_{1,ij}$ is coupled to  currents $J_{2,ij}=XJ_{1,ij}$ along the same bonds in the layer 2. A typical energy of this combined element is $\delta E_{12} \approx T/2 \approx 0.25$.     
This strong asymmetry between the layers has immediate implication for the windings along $z$-direction -- the minimal length $M$ of the element $J_i$ must be $M=2$ in Eq.(\ref{GenM}). Thus, the stiffness $u_r$ in the limit $u<<1$  becomes
\beq
u_r = \frac{4}{2+ \exp(1/u_V)} \approx 4{\rm e}^{-1/u_V}= u^2, 
\label{UR3}
\eeq   
where the asymptotic expression $ u_V=\frac{1}{2 \ln (2/u)}$ \cite{Villain,Kleinert} has been used.
The results of the simulations is shown in Fig.\ref{figN2}.
The first striking feature to notice is that $u_r$ does not depend on the layers size $L$ over 2 orders of magnitude of $L$ and over 7 orders of  $u_V$ (which is actually $\sim \ln(1/u)$ of the bare coupling). Second, the numerically found value $u_r$ follows the analytical result (\ref{UR3}) with high accuracy -- even for values $u_V \sim 1$.   Both features are in the striking conflict with the RG prediction stating that $u_r$ should decay as $\propto L^{2(1-T/T_d)}\approx L^{-1.28} \to 0$ in the SP regime ($T>T_d$). 

It should be also noted that the stiffness along the layers (\ref{stif22}) remains finite and much larger than $u_r$, that is,
$\rho=32.3 \pm 0.1$ for all simulated sizes from $L=8$ to $L=960$. This justifies the validity of Eq.(\ref{UR3}) even in the case $u_V \sim 1$.

\section{Extending the two-layer model to arbitrary number $N_z/2$  of pairs of layers}\label{Sec:Nz}
As it became evident from the previous analysis, no SP can occur in the double layer. Referring to the sketch of the possibilities, Fig.~\ref{PD}, the option (b) is realized. Here we will address a possibility
of SP in a $N_z$-layers setup. In other words, we will be looking for a behavior where the renormalized stiffness (the inter-layer Josephson coupling) $u_r$ decays as a function of $N_z$ in the limit $L \to \infty$, while the stiffness along planes remains finite.[This would be a "weaker" version of the SP].
  
\subsection{RG solution}\label{sec:RGNz}
We consider the PBC setup: the odd $z=1,3,5,7,...$ and the even $z=2,4,6,..$ layers are characterized by the inplane stiffnesses $K_{11}$ and $K_{22} > K_{11}$, respectively, with the nearest layers coupled by the current-current term $\propto K_{12}$ (the same for all pairs of layers) as well as by the Josephson coupling $- u \sum_{x,y,z}\cos(\phi_{z+1} -\phi_z)$ , where $\psi_z(x,y)=\exp(i\phi_z(x,y))$ is the
XY variable defined on  a site $(x,y)$ belonging to the  layer $z$.  

In the linearized with respect to $\nabla_{ij} \phi_z$ approximation analogous to Eq.(\ref{2N0}) the model becomes
\beq
H_0&=& \sum_{z=1,3,5,...}\{ \sum_{\langle ij\rangle} H_{z;ij}  
\nonumber \\
&-& u \sum_{x,y} [ \cos(\phi_{z+1}-\phi_z) + \cos(\phi_{z-1}-\phi_z))]\}
\label{NNz} 
\eeq
where the summation runs over odd values of $z$ and  the notation 
\beq
H_{z;ij} = \frac{K_{11}}{2} (\nabla_{ij} \phi_z)^2 + \frac{K_{22}}{2} (\nabla_{ij} \phi_{z+1})^2 
\nonumber \\
+ K_{12}\nabla_{ij} \phi_{z}(\nabla_{ij} \phi_{z+1}+\nabla_{ij} \phi_{z-1})
\label{HNNz} 
\eeq
is used. 

If the compact nature of $\phi_z$ is ignored, the one-loop RG flow equation for $u_r$ reads
\beq
\frac{du_r}{d\ln L }=(2- \frac{1}{2S} \langle (\phi_{z+1} - \phi_z)^2 \rangle_S ) u_r
\label{RGN}
\eeq
where $\langle ... \rangle_S$ refers to the RG shell integration over the inplane momenta $\Lambda/(1+S) < |\vec{q}| <\Lambda$,
with $\Lambda \sim 1/L$ being the cutoff and $S \to 0$. We note that, due to the PBC along $z$-direction, the mean $\langle (\phi_{z+1} - \phi_z)^2\rangle$ does not depend on $z$.  

Using discrete Fourier representation along $z$ direction with doubled unit cell containing two layers (the odd and the even) with two sorts of
phases $\phi_z=\phi^{(1)}(z)$ and $\phi_z=\phi^{(2)}(z)$ along odd and even layers, respectively, the part $H_{z;ij}$ can be diagonalized
and the correlator in Eq.(\ref{RGN}) found. This gives Eq.(\ref{RGN}) rewritten as
\beq
\frac{du_r}{d\ln L }=2\left(1- \frac{T}{T_d} \right)u_r, 
\label{RGN2}
\eeq
where 
\beq
T^{-1}_d=\frac{1}{4\pi N_z}\sum^{(Nz/2)-1}_{m=0}\frac{1+Y+4X\cos^2q_m}{Y-4X^2\cos^2q_m}, 
\label{TdNz}
\eeq
and the wavevectors along $z$ take values dictated by the periodic boundary conditions $q_m=4\pi m/N_z,\, m=0,1,2,..., (N_z/2) -1$.  
 Here we use the same notations $T=1/K_{11}, X=K_{12}/K_{11}, Y=K_{22}/K_{11}$  introduced in Sec.\ref{sec:RG}.
Thus, at $T>T_d$ RG predicts irrelevance of $u_r$. 

The upper limit on $T$ can be obtained from the requirement of no free 2D vortices in the limit $u=0$. 
There are, actually, two options: i) looking for a composite vortex characterized by phase windings $q_1$ and $q_2$ in odd and even layers, respectively, forming a string of length $N_z$ perpendicular to the layers; ii) considering independent vortices $q_1=\pm 1$ only in odd layers (characterized by smallest stiffness $K_{11}$) and characterized by the temperature $T_{(1,0)}= \pi/2$. As the analysis shows, the option when composite vortices form finite strings (say, $q_2=1$ in layers $z=1,3$ and $q_1=-2X$ in the layer $z=2$ by analogy with the $N_z=2$ case) costs larger energy than in the case ii). The option i) is characterized by the vortex energy $E_v= \pi K_{11} (N_z/2)[(q_1 +2Xq_2)^2 + (Y-4X^2)q_2^2]\propto N_z$. The minimum for a $2X$ integer is achieved at $q_1=-2Xq_2,\, q_2=\pm 1$. Thus, in order to compensate for the factor $N_z >>1$, the system must be very close to the instability $0<Y-4X^2 <2/N_z$.  Simulations in this region turn out to be problematic. Thus, we will conduct simulations in the range $T_d<T<T_{(1,0)}$, provided the condition $T_d <T_{(1,0)}$ holds in the limit where $\delta=Y-4X^2$ remains finite for $N_z>>1$. 
Specifically, the condition $T_d <\pi/2$ reads 
\beq
 \frac{1}{N_z}\sum^{Nz/2 -1}_{m=0} \frac{1+Y + 4 X \cos^2 q_m}{Y -4 X^2 \cos^2 q_m} > 8.
\label{TdT}
\eeq
It can surely be achieved for large enough $X$ in the limit $ N_z >>1$. Replacing the summation by integration in this limit and considering $\delta <<1$, Eq.(\ref{TdT}) gives $ \delta < (X /4\sqrt{2})^2$. For the simulations we have chosen $\delta =0.3$
and $X=6$, which gives $T_d \approx 0.983$ with $T=  1.28$ chosen in the middle of the interval between $T_{(1,0)}=\pi/2\approx 1.57$ and $T_d$.  The chosen value of $T_d$ corresponds to the limit $N_z \to \infty$, and for any finite
$N_z$, the actual $T_d$ from Eq.(\ref{TdNz}) is below this value. 

At this   point we note that for any finite $u$ the system is 3D and, strictly, speaking the notion of the BKT transition becomes inadequate for large enough $N_z$: even at $T>\pi/2$ the odd layers would still have XY order due to the proximity to even layers. Here, however, we presume that SP scenario holds and $u_r$ vanishes at large $L$. Practically, simulations are performed at finite $u$ and we always control that the XY stiffness (helicity modulus) along the layers remains finite and independent of $L$.

\subsection{Dual formulation}
The dual formulation in terms of the closed loops of integer J-currents (along bonds in and between the layers) can be achieved similarly to the case $N_z=2$ by reinstating the compactness of $\phi_z$ in Eqs.(\ref{NNz},\ref{HNNz})  through the Villain approach: $\nabla_{ij} \phi_z \to (\nabla_{ij} \phi_z + 2\pi m_{z,ij})$ along the planes and $-u\cos(\phi_{z+1} - \phi_z) \to (u_V/2)(\phi_{z+1} - \phi_z + 2\pi m_{i,z})$ for Josephson coupling,
where $m_{z,ij}$ refers to an arbitrary (oriented) integer defined on the bond $ij$ belonging to the plane $z$ and $m_{z,i}$ stands for an integer on a bond connecting site $i$ in the plane $z$ to the same site in the plane $z+1$. The partition function is obtained as a result of integration over all $\phi_z(i)$ and summations over all bond integers.   
    
The J-currents enter through the Poisson identity  
 $\sum_{m=0,\pm 1, \pm 2,..} f(m) \equiv \sum_{J=0,\pm 1, \pm 2,..} \int dx  \exp(2\pi \i J x)f(x)$ applied to each bond integer.  This allows explicit integration over all phases $\phi_z$ as well as over the bond integers $m_{z,ij}, m_{i,z}$. 
There are two types of J-currents: inplane $J^{(a)}_{z,ij},\, a=1,2$ within each "elementary cell" (along $z$)  and between the planes $J_{i,z}$. The label $a=1$ refers to J-current defined on the bond $ij$ belonging to a plane with odd $z$. Accordingly, $J^{(2)}_{z,ij}$ stands for the inplane current on the layer with even $z$. Then, $J_{i,z}$ denotes the current from the site $i$ from the plane $z$ to the plane $z+1$.
The integration over phases $\phi$ generates the Kirchhoff constraint --- similarly to the bilayer case. 

Finally the ensemble can be represented as
\beq
Z&=& \sum_{\{\vec{J}\}} \exp(- H_J ), 
\nonumber \\
 H_J&=&\frac{1}{2} \sum_{ij; z,z'}V_{ab}(z-z') J^{(a)}_{z,ij}J^{(b)}_{z',ij'} 
\nonumber \\
&+& \frac{1}{2 u_V}\sum_{i,z} J^2_{z,i},
\label{Hdual}
\eeq 
 where the matrix $V_{ab}(z-z')$ is defined in terms of the matrix $K_{ab}$.
It reflects the asymmetry between odd and even layers. 
Explicitly,
$V_{11} (z)=YV_{22}(z)$, for $z=z-z'$ being even, describes the interaction between odd layers, and $V_{22}(z)$ is defined between even layers; $V_{12}(z)= - X [V_{22}(z+1) +V_{22}(z-1)]$ refers to the interaction between odd and even layers (that is, $z$ is odd), and
\beq
V_{22}(z)=\frac{2T}{N} \sum_{q_m} \frac{\cos(q_m z)}{Y - 4X^2\cos^2(q_m)},\,
\label{Vzz22}
\eeq  
with $ z=0,\pm 2, \pm 4, ...$ and
the summation running over $q_m=4\pi m/N, m=0,1,..., N/2 -1$.

The dual formulation for $N_z$ layers allows obtaining the asymptotic expression for $u_r$ within the same logic
used for deriving Eq.(\ref{UR3}). We will repeat it here. The loop formation can be viewed as a process of  random walks of
two ends of a broken loop -- exactly along the line of the Worm Algorithm \cite{WA}. Such a walk of each end is controlled by energetics of creating one bond element $|J|=1$ in a randomly chosen direction -- either along a given plane or perpendicular to it. Very similar to the  
case of the two layers, the energy to create such an element alone along the odd layer costs energy $>>T \sim 1$, while the same
element along an even layer costs energy $\sim 1$. The only option for creating a loop in an odd layer is if its energy is compensated
by parallel elements in the even plane.   This feature is caused by the strong current-current interaction $\sim X$.
Thus, if the walk occurs along $z$-direction from some even layer $z$ toward the neighboring odd layer $z+1$, the subsequent move along the odd layer will be too energetically costly so that the walker would either move further toward $z+2$ layer or will go back to the original layer $z$. Thus, the inter-layer elements are characterized by either $J_{i,z}=J_{i,z+1}=\pm 1$ or $J_{i,z}=J_{i,z+1}=0$. The weight of such a process is either $\exp(-1/u_V)$ or $1$, respectively.
Even if the walker makes a step or two along the layer $z+1$ (which is a highly improbable event) and then chooses to go toward the layer $z+2$, the contribution to the partition function will be further reduced exponentially by the energy of the element $J$ along the odd plane. Thus, such processes can be ignored, and we arrive at the conclusion that $u_r$  given by  Eq.(\ref{UR3}) must be valid for arbitrary $N_z$
in the asymptotic limit.   
 
\subsection{Numerical results}
\begin{figure}
\vspace*{-0.5cm}
 \includegraphics[width=1.1 \columnwidth]{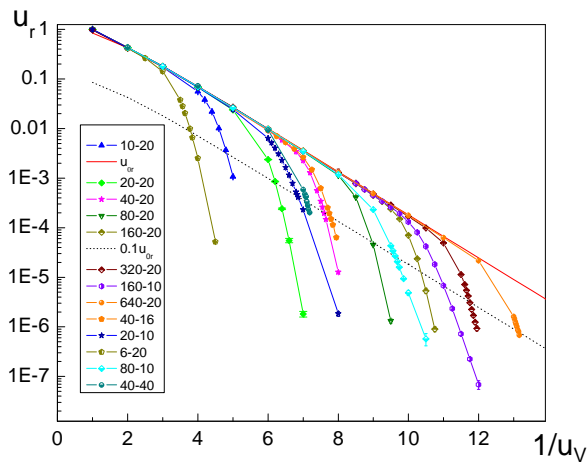}
\vskip-8mm
\caption{(Color online)  Monte Carlo results for the inter-layer stiffness $u_r$ of the model (\ref{Hdual},\ref{Vzz22}) in the SP regime. Dashed orange line is the analytical solution (\ref{UR3}). Dotted black line represents the offset $u_r=0.1$ of the analytical solution (\ref{UR3}). The first and the second numbers in the legend indicate values of $L$ and $N$, respectively}
\label{fig3}
\vskip-5mm
\end{figure}

\begin{figure}
\vspace*{-0.5cm}
 \includegraphics[width=1.1 \columnwidth]{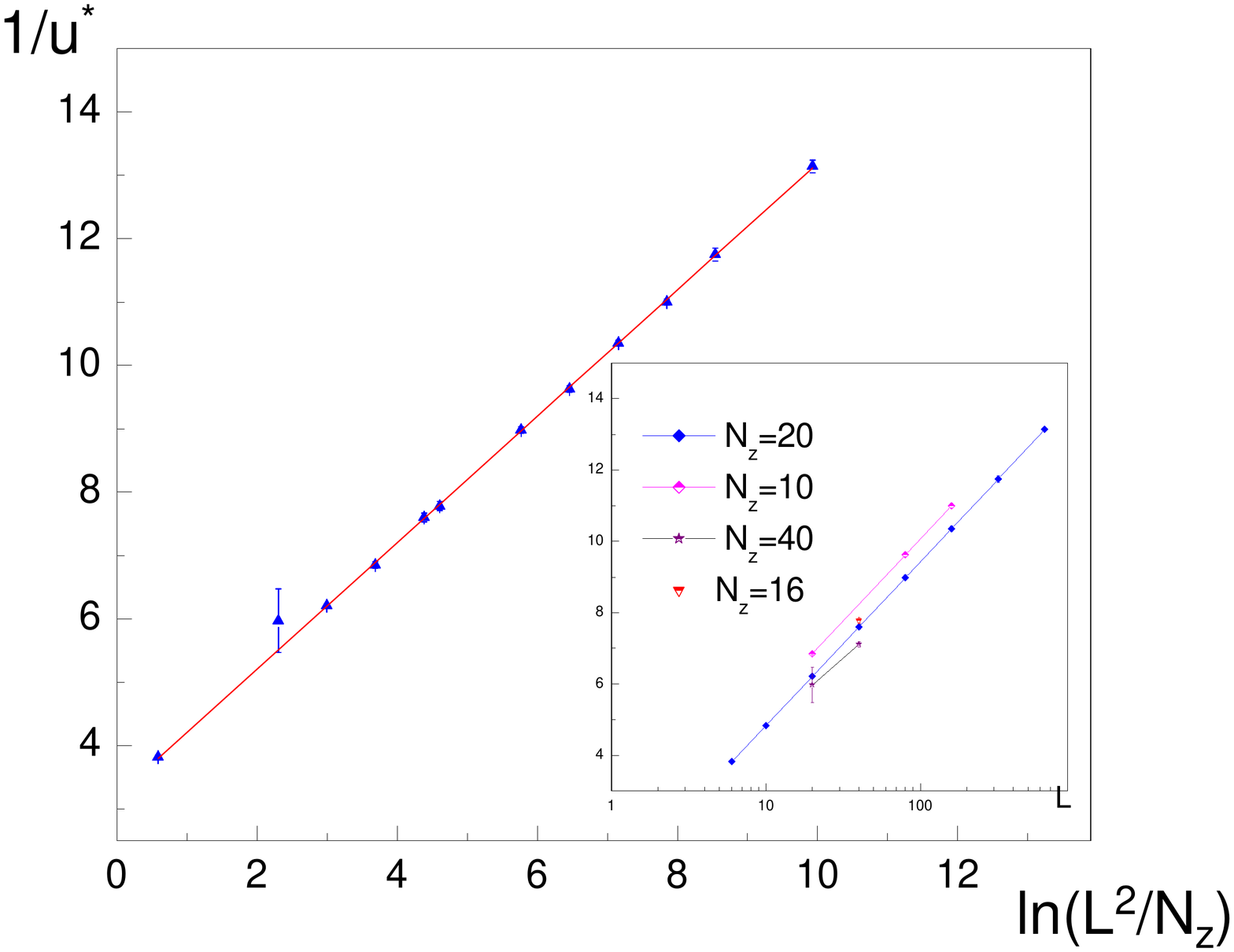}
\vskip-8mm
\caption{(Color online) The values of $u^*$ determined numerically from the data shown in Fig.~\ref{fig3} by finding the crossings of the curves $u_r$ with the offset (dotted) line in Fig.~\ref{fig3}. The linear fit of this line gives the slope $\gamma=1.00 \pm 0.02$.}
\label{fig4}
\vskip-5mm
\end{figure}

The model (\ref{Hdual}) has been simulated by the Worm Algorithm \cite{WA}.
The renormalized inter-layer stiffness $u_r$ was found for a wide range of layer sizes, $6\leq L \leq 640$ and $10 \leq N<\leq 40$. The resulting data is presented  in Figs.~\ref{fig3},\ref{fig4}.  As can be seen in Fig.~\ref{fig3}, the solution (\ref{UR3}) plays the role of the envelop for the family of the curves $u_r$ vs $1/u_V$  for various $L$ and $N$. We note that the stiffness $\rho$ along the layers
(as determined by Eq.(\ref{stif22})) remains independent of the sizes and much larger ($\rho=22.6 \pm 0.5$) than $u_r$. This justifies the applicability of the asymptotic limit for Eq.(\ref{UR3}). We have also controlled that the system is far enough from any possible composite phases \cite{Kaurov} state by measuring the lowest order correlator $\langle \exp(i\phi_z(x,y))\exp(-i\phi_{z'}(x',y')) \rangle$ and observing    
that it exhibits long-range order for values $u\sim 1$ in the limit $N_z\sim L$.[In the composite phase state such a correlator is short ranged]. Thus, the system is well in the putative SP state. Its behavior, however, is drastically different from the RG prediction.

\begin{figure}
\vspace*{-0.5cm}
 \includegraphics[width=1.1 \columnwidth]{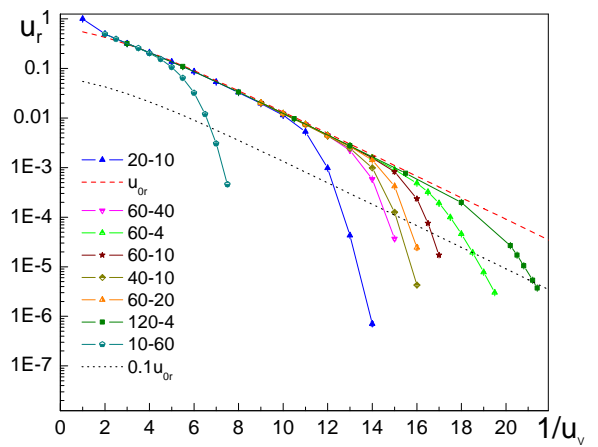}
\vskip-8mm
\caption{(Color online) Monte Carlo results for the inter-layer stiffness $u_r$ of the strongly asymmetric XY model. Dashed orange line is the analytical solution (\ref{urXY}). Dotted black line represents the offset $u_r=0.1$ of the analytical value (\ref{urXY}). The first and the second numbers in the legend indicate values of $L$ and $N$, respectively.}
\label{fig1}
\vskip-5mm
\end{figure}
\begin{figure}
\vspace*{-0.5cm}
 \includegraphics[width=1.1 \columnwidth]{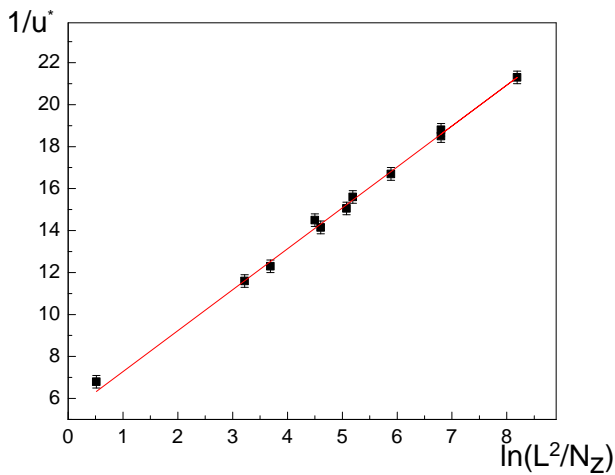}
\vskip-8mm
\caption{(Color online) The values of $u^*$ determined numerically from the data shown in Fig.~\ref{fig1} by finding the crossings of the curves $u_r$ with the offset (dotted) line in Fig.~\ref{fig1}. The linear fit of this line gives the slope $1.95 \pm 0.05$.}
\label{fig2}
\vskip-5mm
\end{figure}

At this point we should discuss the deviations of the numerical curves from the analytical result seen in Fig.\ref{fig3}. 
To some extent this behavior could have been interpreted as the evidence of SP.
There is, however, one important observation: the value of $u_V=u^*$ below which such suppression begins {\it decreases} as   
\beq
(u^*)^{-1}  = \gamma\ln(L^2/N_z), \, \, \gamma =1.00\pm 0.02
\label{UgM2}
\eeq
for $L^2/N_z >>1$ in the main logarithmic approximation. This behavior is demonstrated in Fig.\ref{fig4}, where the value $u^*$ corresponds the offset for $u_r$ being taken at 1/10 of the value given by the analytical expression (\ref{UR3}). The suppression of $u^*\to 0$ in the limit $L\to \infty$ indicates that it is
a finite size effect. In other words, starting from small $L$ at some $u_V <1$, the renormalized stiffness $u_r$ can be well below the asymptotic limit (\ref{UR3}). As $L$ increases while $u_V$ is kept fixed, $u_r$ eventually approaches the  limit (\ref{UR3}). 
This feature is clearly in a stark contrast with the RG prediction of SP where $u_r$ is supposed to flow to zero for fixed $u_V$ as a power of $L \to \infty$. 

The deviation from the limit (\ref{UR3}) has a very simple explanation:  it is essentially a consequence of the generic exponential suppression of any order in a quasi-1D limit $N_z \to \infty$. [There is no such suppression in the case of $N_z=2$ because the loops along z-direction are independent from the inplane ones]. In the context of our system it can be interpreted as the effect of zero modes (in each layer) fluctuations.
Indeed, excitations along the planes renormalize $u_V$ to $u_r$ at short distances. Then, as long as $u_r L^2 << K_{11}$ the only remaining lowest energy degrees of freedoms are zero modes, that is, excitations with $\phi_z=\phi^{(0)}_{z}$ being independent from the $x,y$ positions along the planes (and dependent on $z$). Then, the effective action becomes
\beq
S_0= -u_rL^2 \sum_z[\cos(\phi^{(0)}_{z+1} - \phi^{(0)}_{z})].
\label{zero}
\eeq   
It should be used in the partition function $Z=\int D\phi^{(0)}_z \exp(-S_0)$. Its analytical evaluation gives 
\beq
u^{(0)}_r \sim \exp( -N_z/(u_rL^2)) 
\label{zero}
\eeq   
for the value of the stiffness along $z$-direction -- that is, the renormalised $u_r$ on large scale. Keeping in mind that at short scales $u_r \sim \exp(-1/u_V)$ for $u_V<<1$, one arrives at the relation 
(\ref{UgM2}) with $\gamma=1$. It corresponds to the requirement $N_z/(u_rL^2) \approx 1$ for some $u_V=u^*$ so that Eq.(\ref{zero}) describes the exponential suppression at $u_V <u^*$. 

Clearly, such a quasi-1D suppression is also present in the standard XY model (where no SP are anticipated to exist).
    In order to demonstrate this explicitly we have also simulated a simple XY model given by the system  
\beq
Z_{XY}&=&\int D \phi_z \exp(-H_{XY}),\,\, 
\label{Hxy} \\
H_{XY}&=&-  \sum_{\langle ij \rangle,z} [ \tilde{K}  \cos(\nabla_{ij} \phi_z) + u   \cos(\nabla_{z} \phi_z)], 
\nonumber
\eeq
with some $\tilde{K}>>1$ (guaranteeing that no BKT transition occurs in each layer for $u=0$), and $0<u<<\tilde{K}$.
In the dual representation this system is described by
\beq
H_{XY} \to \tilde{H}_{XY} =  \sum_{\langle ij\rangle,z} \frac{1}{2\tilde{K}} J_{z,ij}^2 + \sum_{i,z}\frac{1}{2u_V} J_{i,z}^2,
\label{Hxy2}
\eeq 
where $J_{ij,z}$ and $J_{i,z}$ are the same J-currents introduced above for the model (\ref{Hdual}). The results of the simulations of this model are presented in Fig.\ref{fig1},\ref{fig2}.
In the asymptotic limit the inter-layer stiffness is described by Eq.(\ref{urXY}). Then, according to the above discussion  the value $u^*$
determining the start of the deviations is given by $(u^*)^{-1} = 2\ln(L^2/N_z)$, that is, with the slope $\gamma=2$ which should be
compared with the numerical value $\gamma=1.95 \pm 0.05$ in Fig.\ref{fig2}.

Thus, both models demonstrate essentially the same 3D behavior. The only difference is the slope of the renormalized Josephson coupling $\ln u_r$ vs its bare value $u_V$. It is determined by the minimal length of the elementary J-current in the direction perpendicular to the layers.

\section{Discussion}\label{sec:dis}
The RG approach to 2D systems proves to be every effective in many cases including 2D XY model when it can be mapped on the Sine-Gordon (SG) one \cite{Polyakov}. A successful implementation of the RG analysis to the Josephson coupling was demonstrated in Ref.\cite{Kane_Fisher} where a single weak link can make one channel Luttinger liquid insulating. 

The merit of RG, however, should be taken with caution when applied to the dimensional reduction situations in layered systems. In this case there is no exact mapping between XY and  SG representations at finite inter-layer Josephson coupling, and the approximation ignoring the compact nature of the variables becomes uncontrolled. As our analysis of one particular layered system shows, no SP exists in such a system despite the RG prediction: the system shows essentially the 3D behavior of the asymmetric XY model.

While it is not obvious to us what is wrong with treating Josephson coupling between 2D layers by RG  in the original representation of fields, the dual formulation in terms of the closed loops gives a very important insight. Specifically, the SP means that as  layer size $L\to \infty$, a suppression of the Josephson coupling between layers would require that the number of times elements of closed loops fluctuate between layers must scale slower than $L^2$ so that the density of such events is zero in the limit $L=\infty$.  The loops statistics, however, is controlled by local energies of creating finite elements and the entropy due to 6 directions in 3D vs 4 along layers. Thus, as long as there is a finite energy to cross between neighboring layers, the entropy will lead to a finite density of crossings for large enough $L$.
Similar argument can be applied to quantum wires in terms of the quantum to classical mapping where imaginary time is treated as an extra dimension. 

The dual approach and the argumentation along the line of the numerical algorithm \cite{WA}, treating closed loops formation as a process of the worm head wondering around and eventually finding its tail,  allowed us to expose the actual stages of the renormalization of the Josephson coupling: i) At small scales Josephson coupling is controlled by exponentially suppressed random and independent (in the asymptotic limit) events of crossings between layers. It can be viewed as an ideal gas of J-currents between the layers. This stage leads to the renormalized coupling, in general, represented by Eq.(\ref{GenM}) with $M=1,2,3,...$.
ii) If the number of layers $N_z$ increases, with $L$ being fixed, quasi 1D fluctuations further suppress the coupling exponentially as demonstrated in Eq.(\ref{zero}). 

Here we have discussed a local model characterized by short range interactions between the inter-layer J-current elements. 
This feature in combination with the low density of such elements justifies the "ideal gas" approximation for them,
which in its turn leads to finite values of the renormalized inter-layer Josephson coupling.

The question may be raised if a  presence of long-range forces  between the inter-plane J-currents $J_i$ can change the situation
and lead to the SP or its weaker version -- where $u_r \to 0$ with the growing number of layers $N_z$ in the limit $L=\infty$.
In this respect we note that in order to realize this, fluctuations of the difference of the J-currents with positive and negative
orientations  must be macroscopically suppressed. In this case the fluctuation of the winding numbers in $z$-direction $\langle W_z^2 \rangle $ will scale slower than $L^2$ so that $ u_r \sim \langle W^2\rangle/L^2 \to 0$. This may be caused by  interactions between the inter-layer J-currents decaying not faster than the second power of their separation. More specifically, the following additional repulsive term
\beq
H_{SP}=\frac{1}{2}\sum_{i,j} U(\vec{x}_i -\vec{x}_j) J_i J_j
\label{SPX}
\eeq 
in the simple XY J-current model (\ref{Hxy2}) with $ U(\vec{x})$ having the long range tail $\sim 1/|\vec{x}|^\sigma$ with $\sigma <2$ will generate
the energy contribution $\sim W^2_z L^{-\sigma}$ in terms of the windings in z-direction. Consequently, the renormalized Josephson coupling (\ref{WU}) would scale as $u_r \sim L^{\sigma -2} \to 0$.  

 In one particular example  long-range forces  are introduced into the inter-layer Josephson in the standard XY model (\ref{Hxy})  by some effective gauge-type term $- u\cos(\nabla_z \phi - g_z A_z) + (\vec{\nabla} A_z)^2$, where $\vec{\nabla} A_z$ refers to the derivatives along the layers of some soft mode $A_z$, with $g_z$ being a parameter. The resulting interaction in the dual form (\ref{SPX}) becomes $U \sim g_z^2 \ln(|\vec{x}|)$ and, thus, it suppresses the inter-layer Josephson as $u_r = N_z \langle W_z^2 \rangle/L^2 \sim  1/(L^2\ln L)$ in the limit $L\to \infty$ for fixed $N_z$.

\subsection{Acknowledgments}
One of us (ABK) acknowledges helpful discussions of the sliding phases with Charles Kane and Boris Svistunov. He thanks   KITPC (the program "Precision Many-body Physics of Strongly correlated Quantum Matter", May-June 2014) and Aspen Center for Physics (the program "Beyond Quasiparticles: New Paradigms for Quantum Fluids", August 2015) for hospitality when this work was in progress.  This work was supported by the National Science Foundation under the grant PHY1314469,
and by a grant for computer time from the CUNY HPCC under NSF Grants CNS-0855217 and CNS - 0958379.

\end{document}